\documentclass[aps,prb,10pt,amsmath,amssymb,twocolumn,showpacs,superscriptaddress,floatfix]{revtex4-1}
\usepackage{ifpdf}\usepackage{hyperref}
 \newif\ifpdf
\ifx\pdfoutput\undefined
   \pdffalse
\else
   \pdfoutput=1
   \pdftrue
\fi
\ifpdf
   \usepackage{graphicx}
   \usepackage{epstopdf}
\else
   \usepackage{graphicx}
\fi
\DeclareMathOperator{\Tr}{Tr}\DeclareMathOperator{\Real}{Re}
\DeclareMathOperator{\EF}{\mathrm{\scriptscriptstyle NP}}\DeclareMathOperator{\TLS}{\mathrm{\scriptscriptstyle TLS}}

\DeclareMathOperator{\NP}{\mathrm{\scriptscriptstyle NP}}

\DeclareMathOperator{\res}{res}

\begin{document}

\title{ Lacunas in the optical force induced by quantum fluctuations of quasicontinuum of multipole plasmons }

\author{E.~S.~Andrianov}
\affiliation{All-Russia Research Institute of Automatics, 22 Sushchevskaya, Moscow 127055, Russia}
\affiliation{Department of Theoretical Physics, Moscow Institute of Physics and Technology, 141700 Moscow, Russia}

\author{N.~M.~Chtchelkatchev}
\affiliation{Department of Physics and Astronomy, California State University Northridge, Northridge, CA 91330}
\affiliation{Department of Theoretical Physics, Moscow Institute of Physics and Technology, 141700 Moscow, Russia}
\affiliation{L.D. Landau Institute for Theoretical Physics, Russian Academy of Sciences,117940 Moscow, Russia}
%\affiliation{Institute for High Pressure Physics, Russian Academy of Science, Moscow, Russia}

\author{A.~A.~Pukhov}
\affiliation{All-Russia Research Institute of Automatics, 22 Sushchevskaya, Moscow 127055, Russia}
\affiliation{Department of Theoretical Physics, Moscow Institute of Physics and Technology, 141700 Moscow, Russia}
\affiliation{Institute for Theoretical and Applied Electromagnetics, 13 Izhorskaya, Moscow 125412, Russia}

\begin{abstract}
We investigate the force between plasmonic nanoparticle and highly excited two-level system (molecule). Usually van der Waals force between nanoscale electrically neutral systems is monotonic and attractive at moderate and larger distances and repulsive at small distances. In our system, the van der Waals force acting on molecule has optical nature. At moderate distances it is attractive as usual but its strength highly increases in a narrow distance ranges (``lacunas''). We show that quantum fluctuations of (quasi)continuum of multipole plasmons of high, nearly infinite degree altogether form effective environment and determine the interaction force while their spectral peculiarities stand behind the large and narrow lacunas in force. We solve exactly the Hamiltonian problem and discuss the role of the dissipation.
\end{abstract}

%05.45.-a Nonlinear dynamics and nonlinear dynamical systems
%72.15.Nj	Collective modes (e.g., in one-dimensional conductors)
%42.50.Ct	Quantum description of interaction of light and matter; related experiments
%42.50.Pq	Cavity quantum electrodynamics; micromasers
%42.50.Nn	Quantum optical phenomena in absorbing, amplifying, dispersive and conducting media; cooperative phenomena in quantum optical systems
%73.20.Mf	Collective excitations (including excitons, polarons, plasmons and other charge-density excitations) (for collective excitations in quantum Hall effects, see 73.43.Lp)
%78.67.-n	Optical properties of low-dimensional, mesoscopic, and nanoscale materials and structures (for magnetic properties of nanostructures, see 75.75.-c; for electronic transport in nanoscale structures, see 73.63.-b; for mechanical properties of nanoscale systems, see 62.25.-g)
%78.67.Pt	Multilayers; superlattices; photonic structures; metamaterials (see also 81.05.Xj, Metamaterials for chiral, bianisotropic and other complex media)
%possible referee A.A. Lisyansky, D. Chigrin, I.E. Protsenko, C. Bruder, , I. Beloborodov
\maketitle

\section{Introduction}
The plasmon resonance is the collective oscillation of electrons in a solid or liquid. Recent progress in understanding plasmon-phenomena at nanoscales have shown that plasmon-assisted Raman-spectroscopy of molecular and biological systems may strongly, by orders of magnitude, increase the resolution and signal strength.~\cite{nie1997probing,campion1998surface,stuart2005biological,brazhe2009new} Plasmon-enhancement effects have been seen in magnetooptics,~\cite{belotelov2007PRL} optoelectronics,~\cite{bergman2003surface,noginov2009demonstration,stockman2010spaser,andrianov2012stationary,Chtchelkatchev2013PhysRevA,rana2011Nature} scanning near-field optical microscopy~\cite{Smolyaninov2005PRL,kawata2007tip} and optomechanics.~\cite{henkel2002radiation,aspelmeyer2012,Restrepo2014PRL}
Apart from optics and spectroscopy there is important question about stresses that plasmons induce on quantum objects nearby. Here we focus on the traditional system for quantum plasmonics: molecule interacting with the plasmonic nanoparticle. Neutral nanoparticles and/or molecules in vacuum attract each other at moderate distances (``van der Waals forces'') and repel each other at close ``atom-size'' range.~\cite{landau1977v3} We show that quantum interaction between a molecule (or a quantum dot) and nanoparticle lead to the origin of deep and sharp attractive wells (lacunas) in the interaction force. There is more or less universal multidisciplinary paradigm, that physical effects related to multipoles of high degree should be likely small. However here not a single multipole but the quasicontinuum of multipoles of nearly infinite degree altogether form effective environment and stand behind the interaction force itself and the nature of its fitches.

We consider below one of the simplest system where the effect of nonmonotonic Van-der-Waals (VW) force can be demonstrated: It consists of nearly spherical metallic nanoparticle (NP) and the two level system (TLS) represented by a molecule or a quantum dot, like in Fig.~\ref{fig00}. We suppose that molecule is excited by interaction with external field or with other molecules.  TLS interacts with the modes of plasmonic nanoparticle through the quantum fluctuations of its dipole moment. Such quantum system, excited molecules and the plasmonic nanostructure, are usual for the near-field microscopes where the the plasmonic nanostructure is placed at needle of the Scanning Plasmon Near-Field Microscope.~\cite{Specht1992PRL} %For simplicity, doing analytical calculations we will focus on the spherical metallic nanoparticle. The case of nanoparticle with general form we discuss below. %Such choice of geometry has an unusual mode density. More concrete eigenfrequencies have condensation point.

The interaction of the TLS and nanoparticle is related to quantum fluctuations of electromagnetic field. So it is natural that the interaction strength appears to be governed by the dimensionless parameter $\alpha$ proportional to the nondiagonal matrix element of  TLS dipole moment, $d_{\rm eg}$, where ``g'' denotes the ground state and ``$e$'' denotes the excited state, see Fig.~\ref{fig00}. The natural normalization energy parameters of the problem in hand are the plasma frequency $\omega_{\rm pl}$ and the TLS level spacing, $\omega_{\TLS}$. Here $\omega_{\rm pl}\sim\omega_{\TLS}$. The natural length unit is the radius $a$ of the nanoparticle. We will show below that $\alpha=\frac{|d_{eg}|^2}{2\omega_{\rm pl}\hbar a^3}$.

We find analytically the quantum state $\Psi(t)$ of the system, TLS$+$NP, and calculate the force acting on the TLS from NP.  The formation of the deep wells in the interaction force $F$ is illustrated in Fig.~\ref{fig00}. The wells form near the interface of the nanoparticle and they are the most pronounced for $\alpha\ll 1$. We show below  that small $\alpha$ is natural for the typical system of TLS and nanoparticle recently investigated experimentally.

Dipole plasmons primarily contribute the VW-force at large distances, $r\gg a$, where $F\sim \alpha\hbar\omega_{\TLS} a^6/r^7$. Extrapolating $F$ naively to moderate distances one would get the estimate for the force, $\sim \alpha\hbar\omega_{\TLS}/a$. However exact calculation including quantum fluctuations of plasmon quasicontinuum gives much larger actual value for the force enhanced by the ``giant'' factor $1/\alpha\gg 1$, so $F\sim \hbar\omega_{\TLS}/a$ as follows from Fig.~\ref{fig00}. Moreover, in the bottom of the dip the force is further enhanced up to the order of magnitude (the width of the dip scales as $\alpha^{1/3}$). Near the plasmon nanoparticle electric field becomes very large.~\cite{klimov_book_nanoplasmonics} This well investigated in dipole approximation. Here we indirectly investigate this electric filed enhancement within plasmon quasicontinuum.

There are many plasmon modes in NP: dipole, quadrupole,...,multipole.~\cite{maier2007plasmonics,klimov_book_nanoplasmonics,gaponenko2010introduction,Delga2014PRL} Typically the dipole mode gives the leading contribution to observables for nanoplasmonic problems. Much rare cases include multipole moments into consideration, see, e.g., Refs.~\onlinecite{Lukyanchuk2006PRL,koenderink2010Optlet,Kivshar2012NJP,Delga2014PRL,andrianov2014spontaneous} and investigations concerning electromagnetic forces between nanostructures.~\cite{xu2002PRL,klimov2009van,Sun2014PhysRevA,Stedman2014PhysRevA} In our problem multipole plasmon modes play the key role. There is dependence of the interaction strength between plasmonic modes and TLS dipole moment on the distance between NP and TLS. Therefore the effective frequency of the system (TLS+NP) oscillation also will depend on the distance. Note that there is a relatively large gap between lower plasmonic modes (e.g. dipole, quadrupole) and it becomes smaller and smaller for higher modes~Fig.~\ref{fig01}.  Then it will be rather natural if the effective frequency coincides with the frequency of these lower modes. In this case  we will have a ``collective resonance'' and the dip in the effective interaction potential.

The paper is organised as follows: In Sec.~\ref{sec1} we define the model and solve the Hamiltonian simplified problem. In Sec.~\ref{Discus} we discuss the applicability of the model, especially the dissipation effects and the possibility of experimental realisation of our predictions. In Appendix we move some technical issues related to the derivation of the force at finite detuning.

\section{Interaction between molecule and multipole plasmonic modes\label{sec1}}

\subsection{Hamiltonian}
\begin{figure}[b]
  \centering
  \includegraphics[width=0.99\columnwidth]{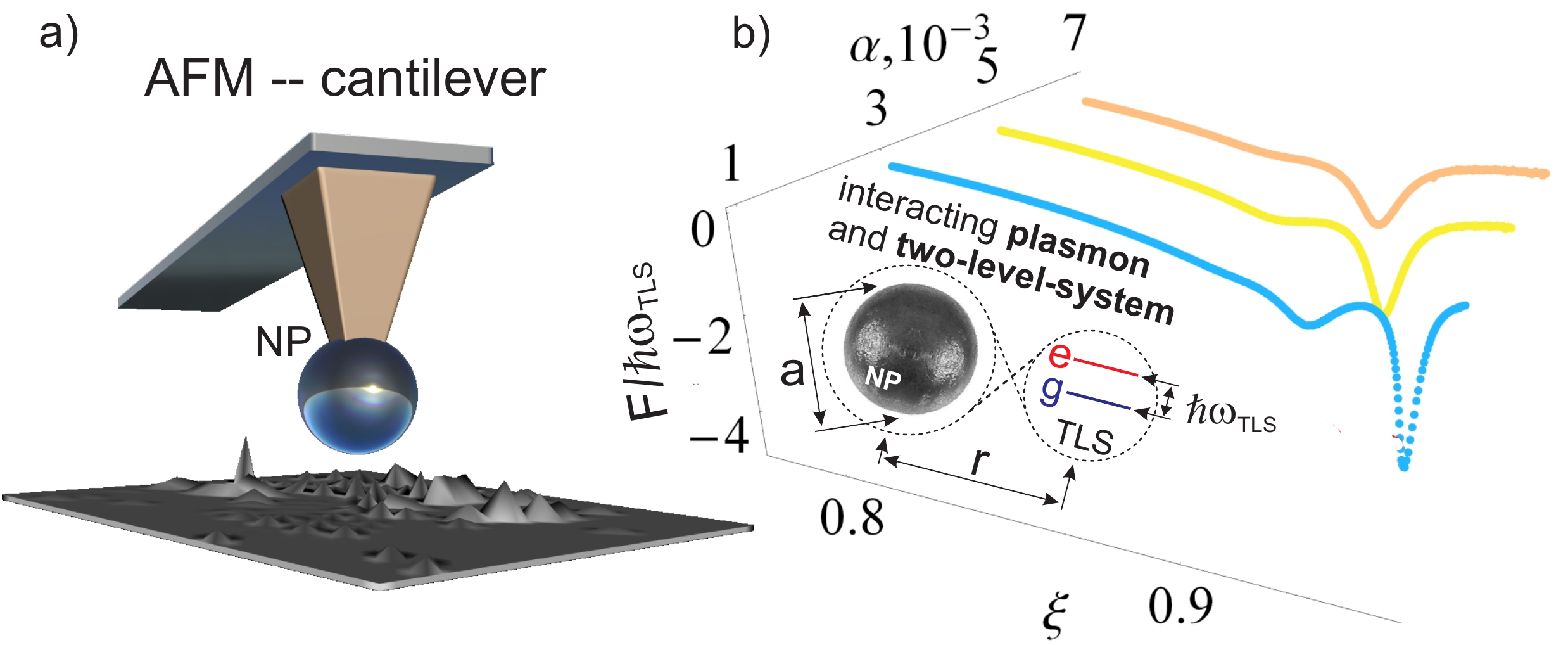}\\
  \caption{(Color online) a) Sketch of plasmonic nanoparticle at the end of the atom force microscope tip scanning molecules at the substrate. b) Van der Waals force between the two-level system and plasmonic nanoparticle as function of the inverse distance $\xi=a/r$. The inset shows the basic characteristics  of the interacting plasmonic nanoparticle and two-level system. }\label{fig00}
\end{figure}
\begin{figure}[h]
  \centering
  \includegraphics[width=0.99\columnwidth]{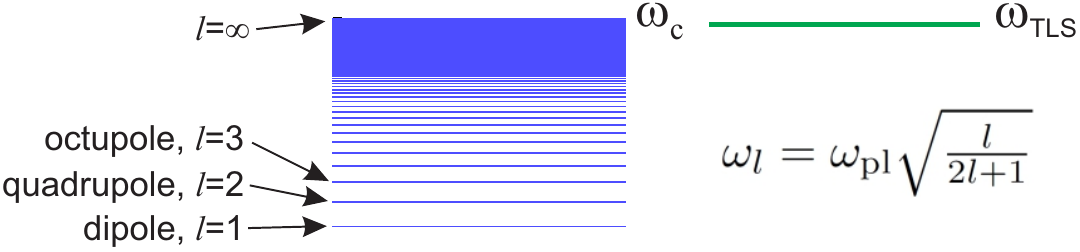}\\
  \caption{(Color online)  The sketch of the relative positions of the plasmon multipole modes, its condensation point $\omega_c$ and the TLS frequency $\omega_{\TLS}$.}\label{fig01}
\end{figure}

Electric field near the nanoparticle can be found generally in the quasistatic approximation using the multipole expansion over the spherical harmonics $Y_{lm}(\varphi,\theta)$,~\cite{voitovich1977} where $\varphi$ and $\theta$ are angles of the spherical coordinates while the integers: $l=0,1,\ldots$ is the order of a spherical function and $-l\leq m\leq l$. Then we get multipole components of electric field for the spherical nanoparticle: ${E}_{lm} =\sqrt{4\pi \hbar \omega_{l} /2a\left(2l+1\right)}$, where  $a$ is the radius of the nanoparticle. Here $\omega_{l}$ is the plasmon resonance frequency in the $l$-th mode.  Within the Drude model: $\omega_{l} =\omega_{\rm pl} \sqrt{\frac{l}{2l+1} }$, where $\omega_{\rm pl}$ is the plasma frequency of the nanoparticle material, see, e.g.,  Ref.~\cite{voitovich1977,klimov_book_nanoplasmonics} for a review. Important property of this expression is the condensation of the plasmon modes~\cite{bergman1992} near the point, $\omega_{c}=\omega_{\rm pl}/\sqrt2$, see Fig~\ref{fig01}. We focuss below on the case when the TLS transition frequency, $\omega_{\TLS}$, falls into the quasicontinuum of the plasmon modes near $\omega_{c}$. It should be noted that the condensation point is present in the plasmon spectrum of nanoparticles with general form.~\cite{footnote1}

Finally, we can write the electric field operator: $\hat{E}_{lm} =-E_{lm} \nabla \varphi _{lm} \left(\hat{a}_{lm} +\hat{a}_{lm}^{\dag} \right)$, where $\hat{a}_{lm}$ is the annihilation operator. So the Hamiltonian for the near electromagnetic field of NP, $\hat{H}_{\EF} =\sum_{lm}\hbar \omega_{l} \left(\hat{{a}}_{lm}^{\dag} \hat{{a}}_{lm} +1/2\right)$.

For the two-level atom:~\cite{Croitoru2007PhysRevLett} $\hat{H}_{\TLS} =\hbar \omega_{\TLS} \hat {\sigma}^{\dag}\hat \sigma$,
where $\hat{{\sigma }}={\left| g \right\rangle} {\left\langle e \right|} $ -- is the transition operator between the excited, ${\left| e \right\rangle} $, and the ground state, ${\left| g \right\rangle} $, see inset in Fig.~\ref{fig1}.

Atom dipole moment $\mathbf{\hat{d}}_{\TLS} =\textbf{d}_{eg} \left[\hat{{\sigma }}(t)+\hat{{\sigma }}^{\dag } (t)\right]$, where $\textbf{d}_{eg} ={\left\langle e \right|} e\mathbf r{\left| g \right\rangle} $. The interaction between the quantum dot and the electromagnetic field, $\hat{V}=-\mathbf{\hat{d}}\cdot\mathbf{\hat{E}}$. In the rotating wave approximation: $\hat{V}=\hbar \sum _{lm}\gamma _{lm}  (\hat{a}^{\dag }_{lm} \hat{\sigma }+\hat{\sigma }^{\dag } \hat{a}_{lm})$, where $\gamma_{lm}$ are the interaction constants.

We focus on the case when the dipole moment is collinear to the line connecting NP and TLS. Then the dipole of TLS interacts only with the symmetric field configurations of NP with $m=0$. So $\gamma_{l,m\neq0}=0$ and
\begin{gather}\label{eq3}
\left|\gamma_{l,m=0} \right|^{2} =\alpha\omega_{\rm pl}^{2}\xi ^{2\left(l+2\right)} {\left(l+1\right)^{2} l^{1/2} }/{\left(2l+1\right)^{1/2}}.
\end{gather}
%Here $\alpha$, $\xi$ and $r$ are defined above.

\subsection{ Admissible parameter range}
Now we discuss the limitations for $\alpha$. For typical quantum dots and plasmonic nanoparticles, $\mu=20$~Debye~\cite{muller2004APL} and $\omega_{\rm pl}=1.370\cdot10^{16}Hz$ for gold particles ($\omega_{\rm pl}=1.366\cdot10^{16}Hz$ for silver ones~\cite{jain2006JPC}). So $\alpha\ll1$. We neglect here radiation to free space. It is valid when the interaction constant between plasmonic modes and TLS is much larger than the radiation rate into free space. E.g. $\gamma_{\rm rad}\ll\frac{\left|\mu_{\TLS}\right|^{2}\omega_{\rm pl} }{2\hbar a^{3}}$. Since $\gamma_{\rm rad}=10^{11}s^{-1}$, this gives the lower limitation: $\alpha\gg10^{-5}$. We discuss parameter values with more details in Sec.~\ref{Discus}.

\subsection{Solution of the Hamiltonian-problem}

The Hamiltonian $\hat{H}=\hat H_{\TLS}+\hat H_{\NP}+\hat V$, where $\omega_l\equiv\omega_{l,m=0}$. We will search the solution of the Shr\"{o}dinger equation with the Hamiltonian $\hat H$ in the form: $\Psi \left(t\right)=A\left(t\right)e^{-i\omega_{_{\TLS}} t}{\left| e,0 \right\rangle} +\sum _{l}B_{l} \left(t\right)e^{-i\omega _{l} t}{\left|g,1_{l} \right\rangle}$ with the initial condition,$\Psi \left(t=0\right)=| e,0 \rangle$. [It is worth mentioning that $\Psi(t)$ corresponds to the entangled plasmon-TLS state except the initial time $t=0$.] Taking into account that $A(t)= \int_{-\infty }^{ \infty }A(\omega)\exp \left(-i\omega t\right)\frac{d\omega}{2\pi}$ we get:
\begin{gather}\label{eqAB}
  A(\omega)=\frac{i}{\omega-\Sigma(\omega)},\qquad\Sigma=\sum_{l>0}\frac{\left|\gamma_{l} \right|^{2} }{\omega-\Delta_{l} +i0 },
\end{gather}
where $\Delta_{l}=\omega_l-\omega_{\TLS}$. Similarly, $B_l(t)=\int_\omega B_l(\omega)e^{-i(\omega-\Delta_l)t}$, and
$B_{l} \left(\omega\right)= \gamma_l A(\omega)/(\omega-\Delta_{l} +i0 )$. For details of these calculations see Appendix~\ref{Appendix1}.

\subsection{Interaction force}
The interaction force between TLS and nanoparticle is equal to $F(r,t)\equiv \langle -\nabla_{r}\hat H \rangle_\Psi$, where $\langle\ldots\rangle_\Psi$ is averaging over $\Psi(t)$. We can either write the force in terms of the eigne energies $E_n$ of the whole Hamiltonian $\hat H$, where $n$ label corresponding quantum numbers: $F(r,t)=\sum_n p_n(r,t)\nabla_r E_n(r)$. Here $p_n(r,t)$ is the probability to occupy the state $|n\rangle$: $p_n=|\langle\Psi|n\rangle|^2$.

The force $F(r,t)$ quickly oscillates at frequencies of the order of $\omega_{\rm pl}$ so we focus on the the time-averaged force. Its graph is shown in Fig.~\ref{fig00}b. The main question is the origin of the sharp dip in $F(r)$. To make progress with the explanation we focus on the average TLS-Hamiltonian, $E_{\TLS}=\langle \hat H_{\TLS}\rangle_{\Psi,t}$, where the additional subscript $t$ means the average also over time. $E_{\TLS}$ has rather simple analytic form, in contrast to the force, and has similar origin nonmonotonic behaviour with distance. So to avoid straightforward but rather cumbersome equations we do the trick: We explain below behaviour of $E_{\TLS}$ with $r$, but all the conclusions apply to the force.

From Eq.~\eqref{eqAB} follows that
\begin{gather}\label{Udef}
  E_{\TLS}/\hbar\omega_{\TLS}=\langle|A(t)|^2\rangle_t=\sum|\res A(\omega)|^2,
\end{gather}
where ``$\res$'' denotes the residue and the sum is take over all the residues. The time averaged perturbation has similar structure, $\langle\hat V\rangle_{\Psi,t}=\sum_l\gamma_l V_l$, where $V_l=2\Real\sum'\res[A(\omega)]\res^* [B_l(\omega)]$. Here $\sum'$ implies the sum of the manifold of equal poles of $A(\omega)$ and $B_l(\omega)$ [this manifold reduces to the poles of $A(\omega)$ as follows from Eq.~\eqref{eqAB}]. Accordingly, the time-averaged force,
\begin{gather}\label{eqF}
  F(r)= \sum_lV_l\nabla_r\gamma_l.
\end{gather}

\begin{figure}[tb]
  \centering
  \includegraphics[width=0.99\columnwidth]{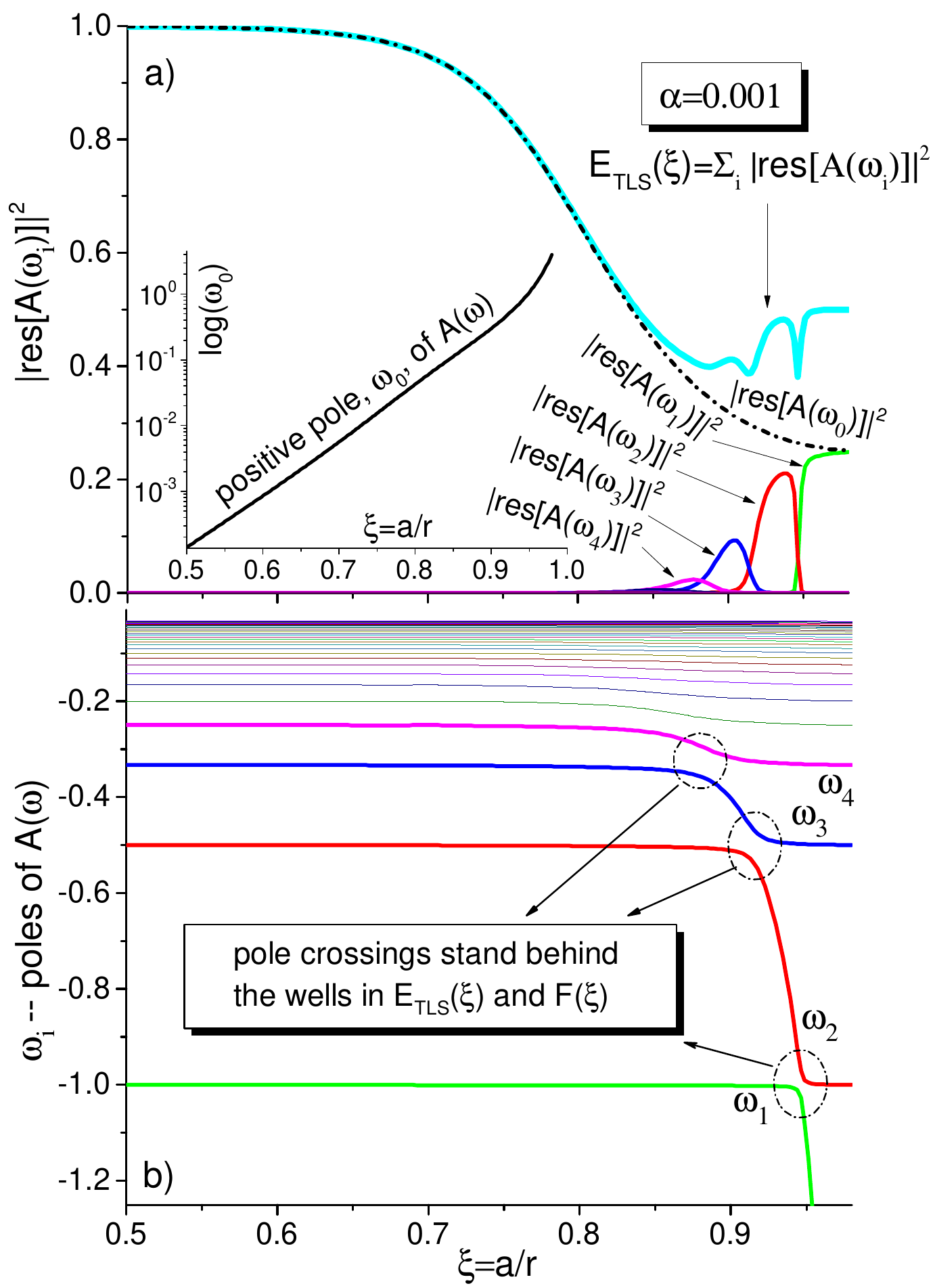}\\
  \caption{(Color online) a) Average TLS-hamiltonian $E_{\TLS}$ is constructed from the residue squares of $A(\omega)$. For $\xi=a/r\to 0$ the positive pole [see the inset] gives the main contribution to $E_{\TLS}$. The wells in $E_{\TLS}$ are induced by the ``anticrossing'' of the negative poles. b) Poles anticrossing: At small $\xi$ negative poles $\omega_l$ are close to $-1/l$ [Here $\omega_{\rm pl}/4\sqrt 2$ normalises the poles]. At certain values of $\xi\lesssim1$ the poles may go very close, nearly touching each other. The smaller $\alpha$, the closer poles. This behaviour of the poles is quite similar to the effect of degenerate level ``repulsion'' in quantum mechanics. Here $\alpha=0.001$. This value corresponds to the typical experimental parameters: $\omega_{\rm pl}\sim 10^{16}\,\mathrm{s}^{-1}$, $d_{eg}\sim 30\, \mathrm{D}$ and $a=5\,\mathrm{nm}$. }\label{fig2}
\end{figure}
First we consider asymptotic behavior of $E_{\TLS}$ on large distance, $\xi=a/r\ll1$. Then the main contribution in the sum of denominator in Eqs.~\eqref{eqAB}-\eqref{Udef} is given by the term with $l=1$, that corresponds to the dipole-dipole interaction. So, the poles of integrand are determined by the equation $\omega=\frac{\left|\gamma_{l=1} \right|^{2}}{\omega-\Delta_{l=1}}$, which has the following roots: $\omega_{0}=-\gamma_{l=1}^{2}/\Delta_{l=1}$ and $\omega_{1}=\Delta_{l=1}+\gamma_{l=1}^{2}/\Delta_{l=1}$.  The positive pole $\omega_0$ gives the main contribution, see Fig.~\ref{fig2}a for illustration. Then for TLS-energy we obtain: $E_{\TLS}/\hbar\omega_{\TLS}\approx |\res A(\omega_0)|^2\approx1-2\gamma_{l=1}^{2}/\Delta_{l=1}^2$. Here unity, the first term in $E_{\TLS}$, corresponds to the average energy of the free TLS (in units of $\hbar\omega_{\TLS}$). The second contribution comes from the interaction with NP. Doing similar with the force~\eqref{eqF} we find that at large distance
\begin{gather}%\label{}
  F(r)= -\frac D{r^7},\quad D\approx 6 a^6\gamma_{l=1}^2(\xi=1)/\Delta_1\sim \alpha a^6\hbar\omega_{\rm pl}.
\end{gather}
Dispersion attraction (the London force) includes the interaction between the instantaneous and induced dipoles. The energy of this interaction is inversely proportional to the sixth power of the distance between the dipoles. In our case TLS dipole induces plasmon dipole.

Calculating numerically the force, see Fig.~\ref{fig00}, we assumed for simplicity that the transition frequency of the two-level atom coincides with the condensation point of the metal nanoparticle resonance frequencies, $\omega_{\TLS} = \omega_c=\omega_{\rm pl} /\sqrt{2}$, so $\Delta_{l}\approx-\omega_{\rm pl}/(4\sqrt{2}l)$ [we did the expansion over $1/2l$]. However our conclusions remain qualitatively valid when $|\omega_{\TLS}-\omega_c|\lesssim \omega_{\rm pl}$ as we discuss below in Sec.~\ref{Discus}. This is so since the singular behaviour of $\Sigma(\omega)$ near $\omega=\Delta_l$ makes the structure of $A(\omega)$ poles robust with the respect to the choice of $\omega_{\TLS}-\omega_c$.

\begin{figure}[bt]
  \centering
  \includegraphics[width=0.99\columnwidth]{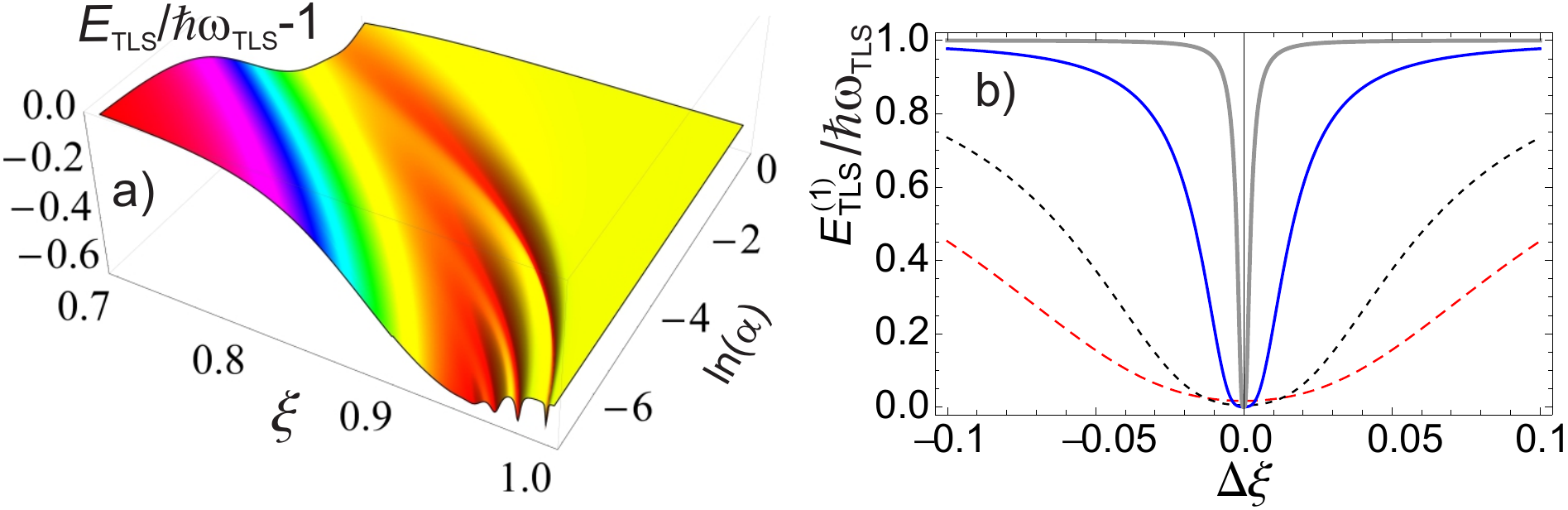}\\
  \caption{(Color online) a) $E_{\TLS}/\hbar\omega_{\TLS}-1$ as the function of $\xi=a/r$ and $\ln(\alpha)$. If $r\gg a$ then $E_{\TLS}$ decays to zero as $-1/r^6$. While $r\gtrsim a$, there are a number of wells in $E_{\TLS}(\xi)$ mediated by the quantum interaction of TLS with plasmons. b) Plot of the analytic approximation~\eqref{eqwell} for $E_{\TLS}^{(1)}/\hbar\omega_{\TLS}$, $\alpha=\{0.001,0.01,0.05,0.1\}$ (from top to bottom).}\label{fig1}
\end{figure}

\subsection{Nature of lacunas in $F(r)$ and $E_{\TLS}(r)$}
As follows from Figs.~\ref{fig1} and \ref{fig2}a, $E_{\TLS}(\xi=a/r)$ becomes very nonmonotonic at $\alpha\ll 1$ and $\xi\lesssim1$. The lacunas in $E_{\TLS}$ become more and more pronounced when $\alpha$ decreases. We remind that the same applies for the force. Below we investigate the origin of the lacunas.

Numerical calculations, see Fig.~\ref{fig2}, show that the lacunas in $E_{\TLS}(\xi=a/r)$ originate from the anticrossing  of the poles at certain $\xi$-range. For instance, the first well corresponds to the anticrossing of the two lowest negative poles, $\omega_1$ and $\omega_2$ of $A(\omega)$, see Eq.~\eqref{eqAB}. The negative poles cross near $\omega=-1/l$, where $l=1,2,\ldots$. [Here and below we choose $\omega_{\rm pl}/4\sqrt 2$ as the unit of the poles.] This is consistent with the structure of $\Sigma(\omega,\xi)$. Our main target are the poles, $\omega_1$ and $\omega_2$, near the first anticrossing situated at $\omega\approx-1$. We distinguish in $\Sigma(\omega,\xi)$ two contributions: the most singular term and the smooth one, $f(\xi)$, coming from infinite set of multipoles:
\begin{gather}
  \Sigma(\omega,\xi)\approx \frac{\alpha\xi^6}{\omega+1}+f(\xi),
  \\
  f(\xi)=\alpha\xi^4\sum_{l=2}^\infty \frac{l^2\xi^{2i}}{-1+\frac1l}.
\end{gather}
The poles are the roots of, $\omega-\Sigma(\omega,\xi)=0$. Now we can write equation for $\omega_{1,2}$ approximately valid near the pole anticrossing:
\begin{gather}\label{eqw12}
  [\omega+f(\xi)](\omega+1)-\alpha\xi^6=0.
\end{gather}
Note that $\alpha\ll 1$ so the last term is the perturbation.

The sum in the definition of $f$ can be evaluated for $\xi \to 1$ and we find up to constants of the order of one,
\begin{gather}\label{eqf}
  f(\xi)\approx{\alpha}/{ \left(1-\xi\right)^3}.
\end{gather}
At certain $\xi_0$ slightly below $1$, $f(\xi_0)=1$. We solve~\eqref{eqw12} near $\xi=\xi_0$. Then $f(\xi)\approx 1+\Delta\xi 2\beta$, where $2\beta=\partial_\xi f(\xi)|_{\xi=\xi_0}$ and $\Delta\xi=\xi-\xi_0$. Then $\omega=-1+\Delta\omega$. Solving Eq.~\eqref{eqw12} for $\Delta\omega$ we get $\Delta\omega_{1,2}= \beta\Delta\xi\pm\sqrt{(\beta\Delta\xi)^2+\alpha^2}$.

Finally we find $E_{\TLS}^{(1)}$ that defines the main contribution to $\sum_{s=1,2}A_{\omega_s}^2$ near the first lacuna in $E_{\TLS}$:
\begin{gather}\label{eqwell}
  E_{\TLS}^{(1)}=\sum_{\sigma=1,2}\left(1+\alpha\xi_0^6/(\Delta\omega_{\sigma})^2\right)^{-2}.
\end{gather}
Eq.~\eqref{eqwell} reproduces the lacuna shape if we plot it as the function of $\Delta\xi$. The width of the lacuna is of the order of $1/\beta\sim (1-\xi_0)^4/\alpha$. It follows from Eq.~\eqref{eqf} that $1-\xi_0\sim\alpha^{1/3}$. So we find for the lacuna-width: $\alpha^{1/3}$. These estimates agree with numerical calculations. We should emphasize that the key role deriving Eq.~\eqref{eqwell} played \textit{the continuum of multipoles} encoded in $f(\xi)$. The same conclusion about the lacunas applies for the force.

\section{Discussion\label{Discus}}
\begin{figure}[t]
  \centering
  \includegraphics[width=0.7\columnwidth]{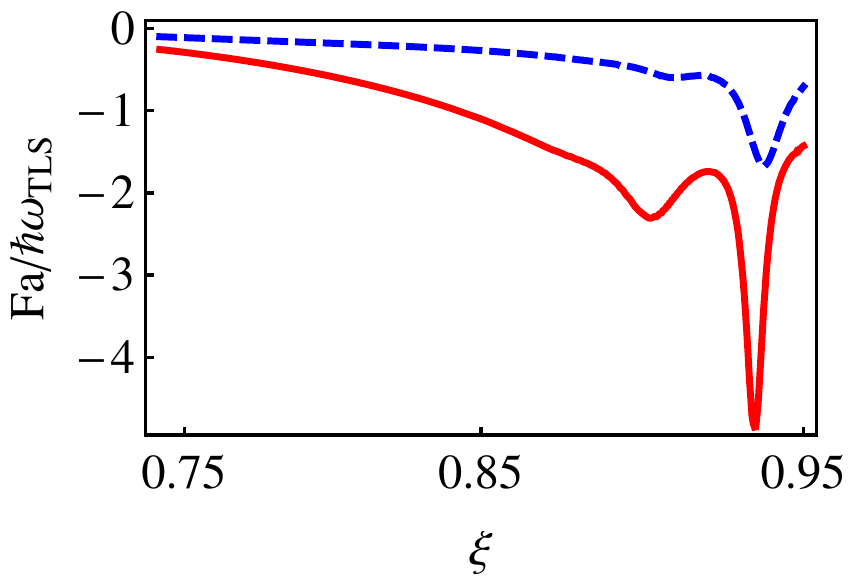}\\
  \caption{(Color online) The mean value of the force (integrated over time and divided by integration interval) acting between NP and TLS. The force is averaged over time up to  $t=10^{-13}s$ (red solid line) and up to $t=10^{-12}s$ (blue dashed line). The difference between the curves is related to the damping that we keep finite here.}\label{figSuppl_dump}
\end{figure}

\begin{figure*}[hbtp]
\centering
\includegraphics[scale=1]{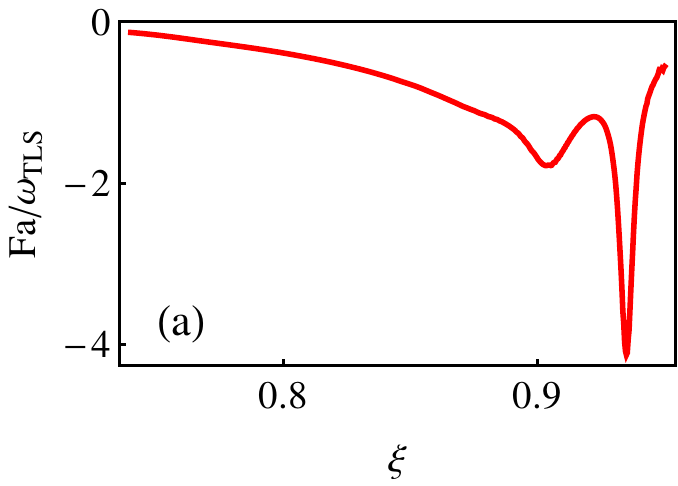}\qquad\qquad
\includegraphics[scale=1]{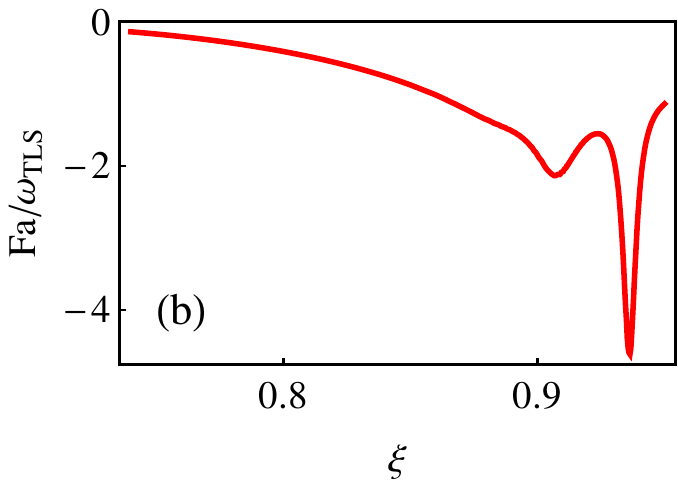}
\\
\includegraphics[scale=1]{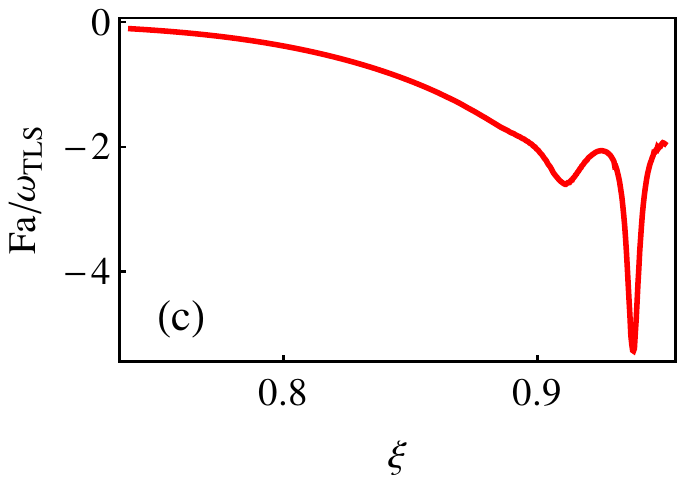}\qquad\qquad
\includegraphics[scale=1]{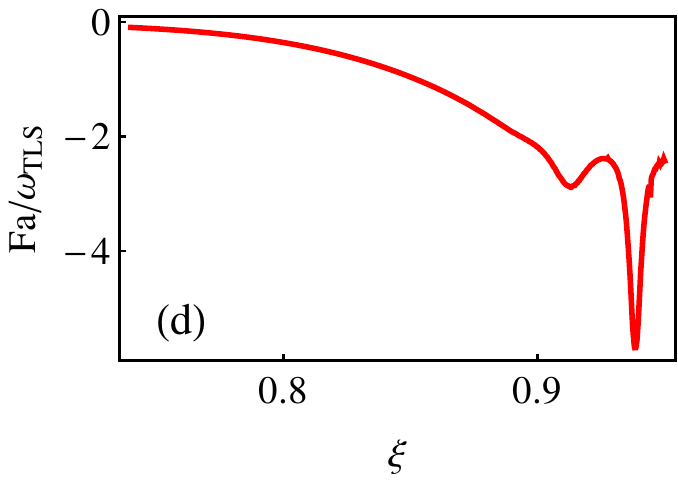}
\caption{(Color online) The dependence of the force on the inverse distance $\xi=a/r$ with different detunings between TLS and condensation point frequencies. a) $\Delta=-5\cdot10^{-2} \omega_{c}$  b) $\Delta=-2.5\cdot 10^{-2} \omega_{c}$ c) $\Delta=2.5\cdot 10^{-2} \omega_{c}$ d) $\Delta=5\cdot 10^{-2} \omega_{c}$. The lacunas in the force are rather stable to detuning.\label{figforces}}
\end{figure*}

\subsection{Parameter estimates}
We estimate the value of the optical force. For optical frequencies~$\gtrsim 10^{15}\,\mathrm{s}^{-1}$ and NP with
radius $a\sim30\,\mathrm{nm}$ we have  $F\sim \hbar\omega_{\TLS}/a\sim10 \,\mathrm{pN}$. This is of the same order or even slightly larger than in typical optomechanical experiment.~\cite{Kohoutek2011NanoLett} We can take much smaller nanoparticle with $a\gtrsim5\,\mathrm{nm}$ and obtain much larger force.

\subsection{The poles, $\omega_l$}
The poles $\omega_l$ (up to the constant) are the energy levels of $\hat H$. In the limiting case $r\gg a$, $\omega_l$ coincide with the eigne levels of $\hat H_0=\hat H_{\TLS}+\hat H_{\NP}$ corresponding to the states $|e,n_l\rangle$, $n_l=0,1$, while at finite $r$ the equation $\omega=\Sigma(\omega)$ reproduces the standard results of the perturbation theory over $\hat V$. The sharp nonmonotonic behaviour of the force ($E_{\TLS}$) with the distance we have got due to the nearly degenerate levels that ``repel'' each-other.~\cite{landau1977v3}

\subsection{Dissipation effects}

We implied above that the system, TLS and NP, is closed. In practise this is not of cause so. Due to quenching effect~\cite{Larkin2004PRB} the damping of the TLS is determined by Joule and radiative losses of excited plasmons in metallic nanoparticle rather than the irradiation into the free space. Due to losses in metal NP surface plasmon lifetime  is short. E.g., in a gold NP surface plasmon lives about $10\,\mathrm{fs}$.~\cite{Klar1998PRL} But these strong radiation losses take place only for large (r>25nm) particles in dipole mode. However oscillations in our case are connected mostly with the condensation point rather than the dipole mode. So losses in high multipole modes are related with the Joule losses in metal which is at least of the order of magnitude smaller than radiation losses. Then, roughly speaking, we should write in the definition of $\Sigma(\omega)$, see Eq.~\eqref{eqAB}, $i/\tau_l$ instead of $i0$, where $\tau_l$ is the characteristic mode life time.

The heat losses in higher multipole plasmonic mode is $10^{12}\,\mathrm{s}^{-1}$ while the radiative loss in the dipole mode $10^{13}\,\mathrm{s}^{-1}$. That estimate gives us the order of magnitude for $\tau_{l=1}$ and $\tau_{l>1}$. Another important point: The force oscillation frequency is of the order of $~\sim 10^{14}\,\mathrm{s}^{-1}$. It should be also noted that the shape of the force dependence is determined by the characteristic value of the poles repulsion which is the order of $\alpha \omega_{c} \approx 10^{14}s^{-1}$. So we may use Hamiltonian without dissipation on short enough time scales. Of course on large time scales our description is not valid. But if there is some pumping acting on quantum dot or electric pumping of nanoparticle with the rate of $10^{-12}\, \mathrm{s}$ we will have the force oscillations during long enough time. Taking into account these estimates we can numerically evaluate the force taking into account the damping times in $\Sigma(\omega)$, see Fig.~\ref{figSuppl_dump}. We see that the lacunas in the force survive while damping is finite.

It should be noted that formally our approach to take damping introducing finite $\tau_l$ in $\Sigma(\omega)$ is insufficient. We should, frankly speaking, either add in addition the fluctuating in time quantum sources simulating thermostat into the Schr\"{o}dinger equation or switch to the density matrix description of our system solving, e.g., Lindblad equations. These program we leave for the forcecoming paper. However our experience with similar models with damping shows that if the effect survives the damping time approximation like we did it would most probably survive within more refined calculations (in the same parameter range).

\subsection{Frequency detuning: nonzero $\omega_{\TLS}-\omega_{c}$}

We investigated above the force in the case of zero detuning, $\omega_{\TLS}-\omega_{c}=0$. Now we discuss the influence of the uncoincidence of TLS transition frequency and NP condensation frequency on the force acting on the TLS. So, now $\omega_{\TLS}-\omega_{c}$ is nonzero.

\begin{figure}[t]
  \centering
  \includegraphics[width=0.99\columnwidth]{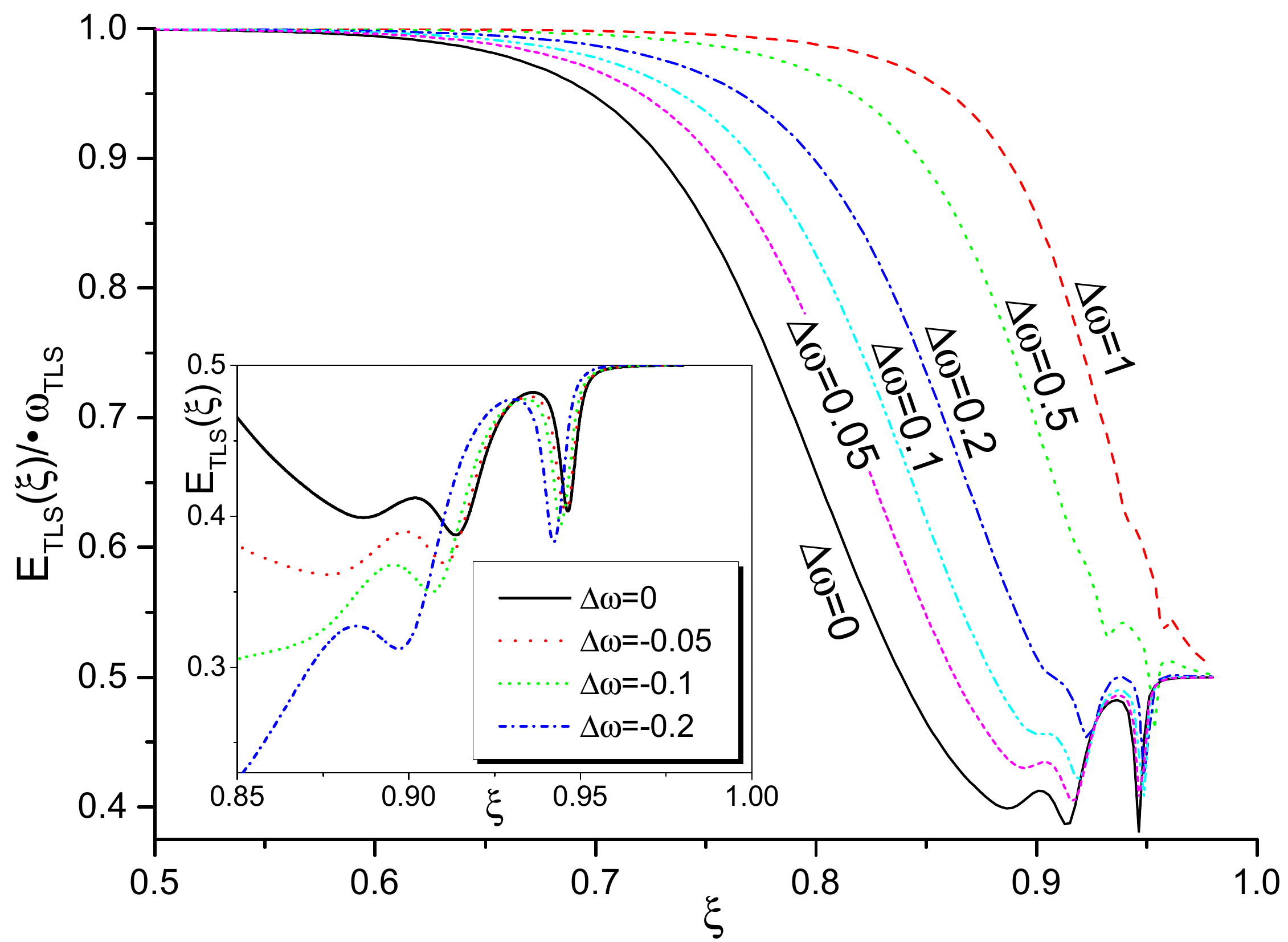}\\
  \caption{(Color online) $E_{\TLS}$ as the function of inverse distance at finite frequency detuning. Here $\Delta\omega=4\sqrt 2(\omega_{\TLS}-\omega_c)\omega_{\rm pl}$. The lacunas in $E_{\TLS}$ are rather stable to detuning.}\label{figSuppl_ETLS}
\end{figure}

We show here that lacunas in the force take place for relatively wide range of the $\omega_{\TLS}-\omega_{c}$. Necessary equations are derived in Appendix. We display in Figs.~\ref{figforces} the force oscillations corresponding to detuning between condensation and two-level system frequencies. In Fig.~\ref{figSuppl_ETLS} we show $E_{\TLS}$. As follows the frequency detuning affects quantitatively on the lacunas however our qualitative picture of lacunas nature remains stable.

\subsection{Entanglement entropy}

\begin{figure}[t]
\centering
\includegraphics[width=0.7\columnwidth]{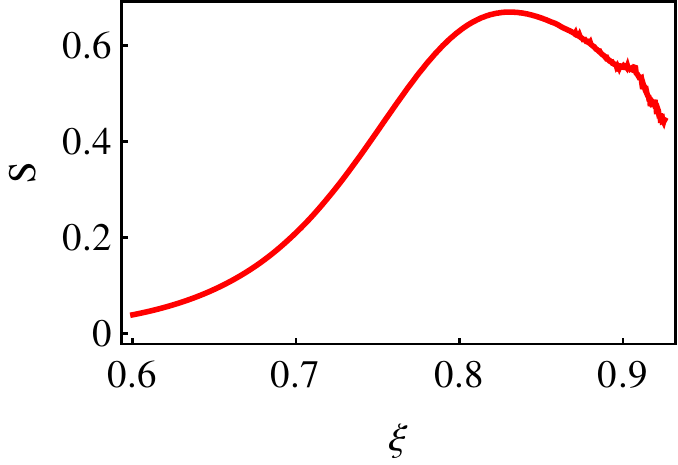}
\caption{(Color online) The dependence of the entanglement entropy on the inverse distance $ \xi=a/r $. \label{figS}}
\end{figure}

It is rather interesting  to investigate the behaviour of the entanglement entropy for NP-TLS system:
\begin{gather}
S=-\Tr(\rho_{\TLS}\log\rho_{\TLS}),
\end{gather}
where $ \rho_{\TLS}=\Tr_{\NP} (\rho) $. In Fig.~\ref{figS} we show the dependence with distance of the entanglement entropy for multipole-NP-TLS system. It follows that entanglement entropy does not have pronounced fitches where the force has. Increase of the entropy at moderated distances can be explained by ``turning on'' more and more multiple plasmonic modes when the distance becomes smaller and smaller. However it is not clear why at small distances where the force shows the lacuna-fitches the entropy decreases. More detailed investigation of this question we leave for the force coming paper.

\section{Conclusions}

The future directions following what we have shown here are interesting and numerous.  The question about the dissipation addressed above requires further clearing up taking into account the $l$-dependence of the relaxation times.  Next, it follows that if the TLS and NP would move with large enough relative velocity $v$ then Landau-Zener transitions between the quasidegenerate levels are expected to contribute the force. So $F=F(v)$. This situation can be relevant in colloids. We considered above only adiabatic regime with  $v\to 0$. Another question that arises is how the force modifies if the plasmon nanoparticle interacts with many molecules.

To conclude, we investigate strong nonmonotonic behaviour of the optical force between the plasmonic nanoparticle and two-level system with distance.  We show that the force strongly grows at moderate distances within short distance intervals. We uncover the nature of the force nonmonotonic behaviour and find that it is mediated by quantum fluctuations and continuum of multipole plasmon resonances of ultra high degree.

\acknowledgments
We thank A. Vinogradov for helpful discussions. The work was funded by RFBR No.~13-02-91177, 13-02-00579, 13-02-00407, NSF Grant DMR 1158666, Dynasty foundation, the Grant of President of Russian Federation for support of Leading Scientific Schools No.~6170.2012.2, RAS presidium and Russian Federal Government programs.

\appendix

\section{The force at arbitrary relation between $\omega_{\TLS}$ and $\omega_{c}$\label{Appendix1}}

\begin{widetext}
We should derive as in the main text the equation which determines the force between NP and TLS at arbitrary detuning. First we remind that the total Hamiltonian is given by
\begin{gather}
\hat{H} =\hbar \omega_{\TLS} \hat {\sigma}^{\dag} \hat \sigma
+\hbar\sum_{l} \omega_{l} \hat {a}^{\dag}_{l} \hat {a}_{l}
+\hbar\sum_{l}\gamma_{l}(\hat{a}^{\dag}_{l} \hat{\sigma}+\hat{\sigma}^{\dag} \hat{a}_{l})
\end{gather}
In this expression the first two terms corresponds to the energy of free TLS and NP-multipole modes respectively, where $ \omega_{\TLS} $ and $\omega_{\NP}$ is the frequencies of TLS and NP respectively. The third term corresponds to the interaction between NP and TLS, here $ \gamma_{l} $ is the interaction strength  between NP multipole mode and TLS, see Eq.~\eqref{eq3}. Here we express it in the form:
\begin{gather}
|\gamma_{l}|^{2}=\dfrac{|d_{\TLS}|^{2}\omega_{pl}}{2\hbar a^{3}} \xi^{2l+4}(l+1)^2\sqrt{\dfrac{l}{2l+1}}.
\end{gather}

The force acting on the TLS is determined by the standard way as $ F=-\langle\psi| \nabla \hat{V} |\psi \rangle$. We look for the solution of the Schr\"{o}dinger equation $ \textit{i}\hbar \dot{\psi} = \hat{H} \psi$ in the form~\cite{andrianov2014spontaneous}
\begin{gather}
\Psi \left(t\right)=A(t)\exp(-i\omega_{\TLS}t)|e,0\rangle + \sum_{l>0} B_{l}(t)\exp(-i\omega_{l}t)|g,1_{l}\rangle
\end{gather}
This leads to the next expressions for A and B
\begin{gather}
A(t)=\int_{-\infty }^{ \infty } \dfrac{\textit{i}\exp(-\textit{i}\omega t)}{\omega -\sum_{l>0}\frac{\left|\gamma_{l} \right|^{2} }{\omega-\Delta_{l} +i0 }} \dfrac{d\omega}{2\pi}, \qquad
B(t)=\int_{-\infty }^{ \infty } \dfrac{\textit{i}\exp(-\textit{i}\omega t) \exp(-\textit{i}\Delta_{l} t)}{(\omega-\Delta_{l})(\omega -\sum_{l>0}\frac{\left|\gamma_{l} \right|^{2} }{\omega-\Delta_{l} +i0 })} \dfrac{d\omega}{2\pi}
\end{gather}
where $ \Delta_{l}=\omega_{pl}\left( \sqrt{\frac{l}{2l+1}}-\frac{1}{\sqrt{2}}\right) + \Delta  $, where $\Delta=\omega_{c}-\omega_{\TLS}$. Becuase of fast oscillating on optical frequencies we shall average the resulting force on time.
After some algebra we obtain the next explicit expression for force acting on TLS
\begin{multline}
F(\xi,\Delta)= \sqrt{\frac{|d_{TLS}|^{2}\hbar \omega_{pl}}{2a^{5}}}\sum_{l>0}(l+2)(l+1)l^{1/4}(2l+1)^{-1/4}\xi^{l+3}
 \\
\int_{0}^{\infty} dt\exp(i(\omega_{\TLS}-\omega_{l}) \int_{-\infty }^{ \infty } \dfrac{\textit{i}\exp(-\textit{i}\omega t)}{\omega -\sum_{l>0}\frac{\left|\gamma_{l} \right|^{2} }{\omega-\omega_{pl}\left( \sqrt{\frac{l}{2l+1}}-\frac{1}{\sqrt{2}}\right) -\Delta +i0 }} \dfrac{d\omega}{2\pi}
 \\
\int_{-\infty }^{ \infty } \dfrac{\textit{i}\exp(-\textit{i}\omega t) \exp(-\textit{i}\Delta_{l} t)}{(\omega-\omega_{pl}\left( \sqrt{\frac{l}{2l+1}}-\frac{1}{\sqrt{2}}\right) -\Delta)(\omega -\sum_{l>0}\frac{\left|\gamma_{l} \right|^{2} }{\omega-\omega_{pl}\left( \sqrt{\frac{l}{2l+1}}-\frac{1}{\sqrt{2}}\right) -\Delta +i0 })} \dfrac{d\omega}{2\pi} + c.c.
\end{multline}
This formula is illustrated in Fig.~\ref{figforces}.
\end{widetext}

\bibliography{refs}

\end{document}